# A SIMULATION FRAMEWORK FOR FAST DESIGN SPACE EXPLORATION OF UNMANNED AIR SYSTEM TRAFFIC MANAGEMENT POLICIES

*Ziyi Zhao, Chen Luo, Jin Zhao, Qinru Qiu, M. Cenk Gursoy, Department of EECS, Syracuse University, Syracuse, NY*

*Carlos Caicedo, School of Information Studies, Syracuse University, Syracuse, NY*

*Franco Basti, Thales Digital Aviation Customer Success and Innovation, Americas*

## Abstract

The number of daily small Unmanned Aircraft Systems (sUAS) operations in uncontrolled low altitude airspace is expected to reach into the millions. UAS Traffic Management (UTM) is an emerging concept aiming at the safe and efficient management of such very dense traffic, but few studies are addressing the policies to accommodate such demand and the required ground infrastructure in suburban or urban environments. Searching for the optimal air traffic management policy is a combinatorial optimization problem with intractable complexity when the number of sUAS and the constraints increases. As the demands on the airspace increase and traffic patterns get complicated, it is difficult to forecast the potential low altitude airspace hotspots and the corresponding ground resource requirements. This work presents a Multi-agent Air Traffic and Resource Usage Simulation (MATRUS) framework that aims for fast evaluation of different air traffic management policies and the relationship between policy, environment and resulting traffic patterns. It can also be used as a tool to decide the resource distribution and launch site location in the planning of a next generation smart city. As a case study, detailed comparisons are provided for the sUAS flight time, conflict ratio, cellular communication resource usage, for a managed (centrally coordinated) and unmanaged (free flight) traffic scenario.

## Introduction

The introduction of unmanned aircraft systems (UAS) into the global airspace system creates both an opportunity and a challenge for the aviation industry as a whole. The traffic demand from new entrants in low-altitude airspace is forecasted to be orders of magnitude far greater than existing commercial aviation. Demands will increase within the existing system but will also be felt especially in large, metropolitan areas. Several forecasts show that package delivery service alone can generate up to 100,000 operations per day in an area as large as San Francisco Bay. Emerging solutions such as UAS traffic management (UTM) will soon be required to manage the increased traffic within the airspace and much more infrastructure including ground-based communications solutions will be needed to help ensure the safety of the platforms, manned aviation and the general public at large. A debate is still ongoing regarding the most secure, effective way to accommodate the demand from the operators and there is a general acceptance that research activities will still need to be conducted to examine how we will cope with this unprecedented growth.

An ideal UTM solution will have the capability to coordinate the launching of sUAS from different launch sites and determine their trajectories to avoid conflicts while considering several other constraints such as arrival deadline, minimum flight energy, and availability of communication resources. Searching for the optimal air traffic management policy is a combinatorial optimization problem with intractable complexity when the number of sUAS and the constraints increases. As the traffic pattern becomes increasingly complex, it is difficult to foresee the potential for conflicts and to estimate the flight time and communication resource requirements. It is these collective challenges that have motivated us to come up with a simulation capability to closely examine how the increased complexities of the airspace system along with the strains on the communications network interact to safely connect the platforms to UTM systems.

Syracuse University, along with a team at Thales Digital Aviation Solutions has created a Multi-agent Air Traffic and Resource Usage Simulation (MATRUS) framework that aims for fast evaluation of different air traffic management policies and the relationship between policy, environment and resulting traffic patterns. The framework is envisioned as a near-real-time resource distribution tool that can enable informed decisions regarding launch site selection as part of the planning of a next-generation smart city. The MATRUS framework is an integrated environment for air traffic simulation, communication resource estimation, data analysis, and traffic animation. At this time, the platform has been used to study sUAS traffic patterns and communication resource usage. However, it can be extended to include other considerations, such as the impact from weather, the need to stay in radar monitoring area, etc., and be used to simulate air traffic scenarios where there are heterogeneous types of sUAS missions being carried out.

In a recently conducted case study, MATRUS incorporated two different traffic scenarios, a point-to-point free fly scenario without any traffic management and a managed scenario with ground centralized traffic management. The traffic management algorithms in MATRUS have the ability to schedule and route air traffic over a metropolitan area that has high sUAS densities by planning each sUAS trajectory in advance. This ensures sUAS safety by proactively avoiding conflicts while maximizing the usage of the communication resources. This comes with a trade-off of marginally extended flight time and increased launching intervals. Detailed comparisons between the scenarios are provided for the sUAS flight time, conflict rate, and the cellular communication resource usage among other variables. This paper will describe the platform, its application and its benefits to those progressive communities seeking to integrate drone operations as part of their smart city planning. It will also show its viability to address challenges in the emerging urban air mobility market.

The rest of the paper is arranged as follows: In Section II, we review related work in UTM. Section III introduces the motivation of this paper. This is followed in Section IV by details about the MATRUS framework. Section V describes our experiments and evaluation of the results. Finally, Section VI summarizes this work.

## Related Works

Recent works from NASA [1][2] identified the need for simulation environments to study the most efficient low altitude airspace organization and the capacity of the low altitude airspace to safely accommodate the high demands derived from commercial use of this airpace by sUAS. Those works do not address the relationship between airspace and the required communication infrastructure capacity to safely and efficiently handle the traffic. This relationship is highly critical to dense beyond-line-of-sight (BVLOS) traffic where the majority of the sUAS operators will potentially use a mobile ground-based communication network to monitor and control the sUAS.

The potential traffic demand derived from sUAS commercial usage and in particular from package delivery have been discussed in [3]. The debate about intrinsic airspace capacity and the question if the free flight concept can safely accommodate very dense traffic have been previously discussed in [4][5][6][7] and remain an open topic both for unmanned and manned aviation.

A very strict and rigid airspace structure to handle dense operation in the urban low altitude environment was suggested by UTM NASA in [8], and this reinforces the view of the authors that more research is required to architect a UTM solution capable of handling such high traffic demand and that in some situations free flight operations with fully decentralized trajectory planning are not feasible or will result in very inefficient airspace operations.

## Motivations for the MATRUS framework

Although analytical models have been investigated to estimate sUAS traffic density and flight time, to the best of our knowledge, few of them have the ability to consider realistic flight scenarios. Furthermore, the existing analytical models lack flexibility required to address our target environment. Most models only take a limited subset of sUAS related information, such as the location of launching/landing area and launching probability, as an input but ignore the constraints from the airspace environment. For example, a typical airspace constraint is the establishment of no-fly-zones, which prohibit the launching, landing and operation of small UASs in a certain geographic area.Due to the lack of flexibility, the analytical approach cannot be easily modified to handle traffic estimation under different airspace structures and management policies. For example, if the airspace is divided into sky-lanes that are either parallel or orthogonal to each other, a completely different approach must be used to analyze the traffic under such structure compared to a fully non-structured airspace and free-flight point-to-point traffic pattern.

The other major limitation of existing simulation environment is the lack of consideration for the command and control (C2) link used during the flight. The availability and reliability of the C2-link is one of key factors maintaining the operation's safety. For this reason, in urban/suburban areas, we anticipate that the communication infrastructure capacity will play a major role in the UTM system. Possible network congestion or outages will lower airspace capacity and as such traffic management policies will be required to dynamically take into account the infrastructure capacity.

As a motivational example, we estimate average sUAS flight time for three different traffic scenarios. In the first scenario, all sUAS follow the simple point-to-point trajectory ("P2P") from the launching area to the landing area. In the second scenario, all sUAS follow the sky-lane flight trajectory approach described in [8]. Since all sky-lane trajectory segments are either horizontal or vertical, we refer to it as "Manhattan" style trajectories. The third scenario is similar to the second scenario, however, a $1800*2700$ m$^2$ no-fly zone was specified in the center. And we refer to it as "Manhattan with restrictions". The analytical model that we used is given in [9].

Table. 1 shows the estimated average flight time and actual (i.e. simulated) flight time of the 3 scenarios. From the table, we can see that the analytical model can generate accurate estimation for the simplest traffic scenario, i.e. P2P. However, it cannot handle the restriction and constraints that we applied on the sUAS trajectory and airspace environment.

**Table 1. Average UAS flight time**

| Traffic scenarios | Analytical | Real |
|---|---|---|
| P2P | 389.16s | 381.95s |
| Manhattan | 387.04s | 487.64s |
| Manhattan w. restrictions | 386.61s | 522.01s |

All aforementioned limitations of the existing analytical models and simulation environments motivated us to develop a simulation based approach that concurrently simulate the low altitude airspace sUAS traffic and the associated C2 link. This allows for the study of both airspace traffic management policies and the relationship between low altitude demand/capacity and ground communication infrastructure demand/capacity.

# The MATRUS platform

The MATRUS platform is an integrated environment for air traffic simulation, communication resource estimation, data analysis, and traffic animation specifically designed to address sUAS traffic in low level altitude airspace. The software components that are part of the platform are shown in Fig. 1. There are three major components: a multi-agent event-driven simulation engine developed over the REPAST (Recursive Porous Agent Simulation Toolkit) Simphony platform, a traffic animation tool implemented using Google Earth API, and a Python based data analysis tool suite. Agent-based modeling is being used to model sUAS' components behavior and to emulate related air traffic phenomena as dynamical systems of interacting agents. The interactions between agents determine the spatial and temporal evolution of a scenario where each independent agent is designed to exhibit individual localized behavior (not globally controlled behaviors). At this time, the platform has been used to study sUAS traffic patterns and communication resource usage from such patterns. However, the simulated scenarios can be modified and allow to support different types of restrictions. and agents (i.e. UAS with different missions: package delivery, surveillance, infrastructure inspection, etc.)

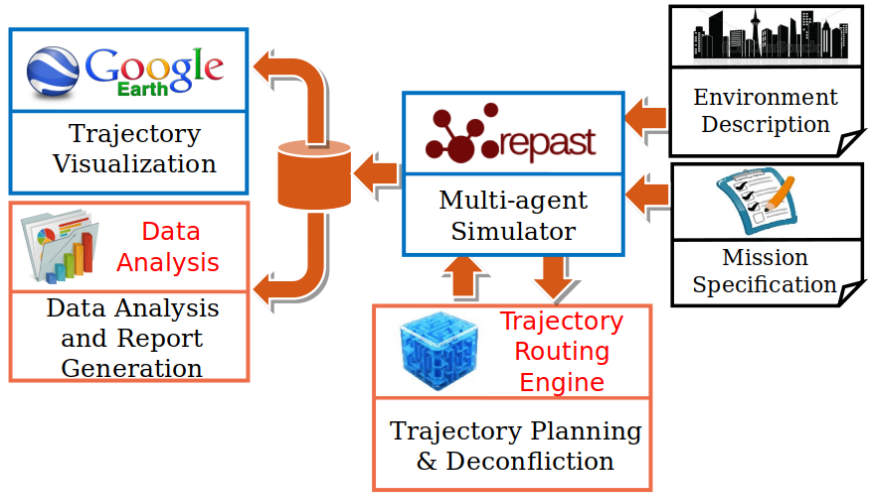

**Figure 1. MATRUS Framework Major Components**

*Agent-based modeling platform*

The core functionality of MATRUS is built on top of the REPAST (Recursive Porous Agent Simulation Toolkit) Simphony platform developed by the Argonne National Laboratory [10]. REPAST provides a set of tools for the development of agent-based models in Java along with data collection, data analysis, and error reporting capabilities. Python based code is used for the generation of visualizations/maps, data processing, and report generation.

The team used agent-based modeling techniques for the modeling of sUASs and their related air traffic phenomena along with the infrastructure components of sUAS as dynamical systems of interacting agents.

The simulator functionalities are composed of several modules. Among these modules, the air traffic module contains the logic to plan, schedule and manages sUAS air traffic. It is composed by sub-modules that carry out tasks such as:

- **sUAS flight scheduling**: This component has the logic to plan the schedule of flight initiation for each UAV agent in a simulation scenario.

- **Trajectory planning:** Logic for specifying the start, end and waypoints of the flight trajectory for one or a group of UAV agents. It will also contain the logic that specifies one or several flight trajectory methods (i.e. direct flight path, restricted path, point-area-point, etc.) and the setting of altitude and speed limits for UAV flights for a particular simulation scenario. By using the integrated centralized routing algorithm, the UAV conflict ratio can be significantly reduced. Moreover, the routing algorithm guarantees the connectivity between each UAV and a cellular base station.

- **Mission specification:** This component has the logic to specify the mission of each UAV agent. Examples of mission types are package delivery, surveillance (hovering), etc. Each mission profile can be enhanced with items such as bandwidth/capacity requirements for data transmission, SINR limits, etc.

Features of the simulator facilitate the study of the interplay between communications network capacity and sUAS traffic. We assume that sUAS will rely on a cellular/5G network with base stations that will support the sUAS-to-ground communications. The communications module in the simulator contains the logic to layout or define the locations and characteristics of the components of the communications network infrastructure that will support the communications between the sUAS and a command & control center. The wireless communication link is characterized considering distance-based path loss models in order to account for the radio frequency (RF) power dissipation as a function of the communication distance. Probabilistic line-of-sight and non-line-of-sight propagation models are incorporated into the link characterization.

*RF Communications and Propagation Model*

The MATRUS framework supports various methods to evaluate the signal strength transmitted from a UAV that arrives at the antenna of a cellular base station. By default, the log-distance path loss propagation model is used. The mathematical expression for this model is given by the following equation:

$$PL_{d_0 \to d}(dB) = PL(d_0) + 10n\log_{10}\left(\frac{d}{d_0}\right) + x,$$
$$d_f \leq d_0 \leq d$$

Where $PL_{d_0>d}$ is the path loss in $dB$ at any arbitrary distance $d$ and $PL(d_0)$ is the path loss in $dB$ at a reference distance $d_0$. The parameter $n$ is the path loss exponent. Typical values of the path loss exponent for various environments are given in Table 2 [11]. The $x$ parameter is a zero-mean Gaussian distributed random variable with standard deviation σ, modeling the shadowing effect [12]. In our framework, the default value of $x$ is set to be 0. In addition, other path loss models can be also integrated into the MATRUS framework, such as Stanford university interim (SUI) model [13][14], Hata model [15], etc.

**Table 2. The Typical Value of Path Loss Exponent**

| Environment | Path Loss Exponent (n) |
| --- | --- |
| Free Space | 2 |
| Urban | 2.7 to 3.5 |
| Shadowed Urban | 3 to 5 |
| Inside a building - Line of Sight | 1.6 to 1.8 |
| Obstructed in Building | 4 to 6 |

*Data Processing*

The data processing module generates statistical information of the flight and resource usage for a scenario or set of scenarios by processing the log files generated by the simulator. One of the key metrics extracted is the UAV density $D$. It is defined as the number of UAVs in a $W \times W$ region across the air space. It is calculated by convolving a $W \times W$ all-pass filter across the entire air space:

$$D(x,y) = \sum_{i,j=0}^{W-1} d(x*S+i, y*S+j)$$

where $d(x, y)$ gives the number of sUAS at location $(x, y)$ and $S$ is the stride size of the convolution. We refer to the matrix $D(x, y)$ as the density map and the matrix $d(x, y)$ as the distribution map. The density map effectively reduces the size of the distribution map by $S^2$ and it gives more smoothly filtered information of the air traffic distribution. By replacing the all-pass filter with a max filter, we can get the maximum density map:

$$D(x,y) = max_{0 \leq i,j \leq W-1} d(x*S+i, y*S+j)$$

The max density map shows the maximum UAV density in the $W \times W$ neighborhood of $(x, y)$. Both density maps and the distribution map are sparse matrices. Hence they are implemented as linked lists to preserve memory efficiency.

## Experimental Results & Analysis

To demonstrate the capability of the MATRUS framework, we present some simulation results that compare the traffic and resource utilization of sUAS under different environmental settings. The environment settings are configured using 6 sets of parameters. One of them specifies the base station configurations; two of them specify the configuration of each single UAV; and the rest of them specify the mission information and flight space configurations. The detailed description of each set of parameters is as follows:

- The configuration of the cellular base station includes the location of each base station ($x_b$, $y_b$) and the number of communication channels ($c$) that are allocated for UAV communications. In most experiments, this parameter is fixed to be 8 channels in the simulation.
- The trajectory type ($T$) for each sUAS: In the experiments, this parameter is set to be either straight (point-to-point) or Manhattan style.
- Routing ($R$) indicates the style of trajectory planning of a UAS. Since there is only one trajectory planning algorithm implemented in the simulator at this moment, $R$ is currently a binary variable.
- sUAS flight mission generation interval ($T_{min}$) indicates the minimum interval for one launching area to generate the launch of a UAV. In the experiments, the number is selected to be 10 (simulation) seconds.
- The configuration of no-fly zones in the simulated airspace is described by the number and location of no-fly zones in the simulation.

In this paper, each combination of parameters defines a specific scenario. The results reported in this section is the average of 10 runs for each scenario. For each single run of one scenario, the simulation time is 20,000 time steps. In the simulation process, each time step corresponds to 1 second. We observed that after around 300 time steps, the simulation behavior stabilized. Therefore, the simulation length is sufficient for us to analyze different scenarios.

Four sUAS launching areas and four landing areas are distributed across a 90 square miles. Their locations are selected based on the distribution of business and residential areas in the Syracuse, NY region. Each launching area has a different launching probability. In a given interval, each launching area will decide whether to launch a UAV

Based on probability for a launched UAV, the system randomly chooses one landing area from four candidate sites. The sum of the probabilities of the four landing areas is equal to 1. The setting of the simulated environment is illustrated in Fig. 2. In the figure, the red boxes represent the UAV launching areas, the purple boxes stand for the UAV landing areas. The locations of 10 base stations are also marked on the map. The coordinates of those base stations are set based on the actual base station facilities registered with the FCC.

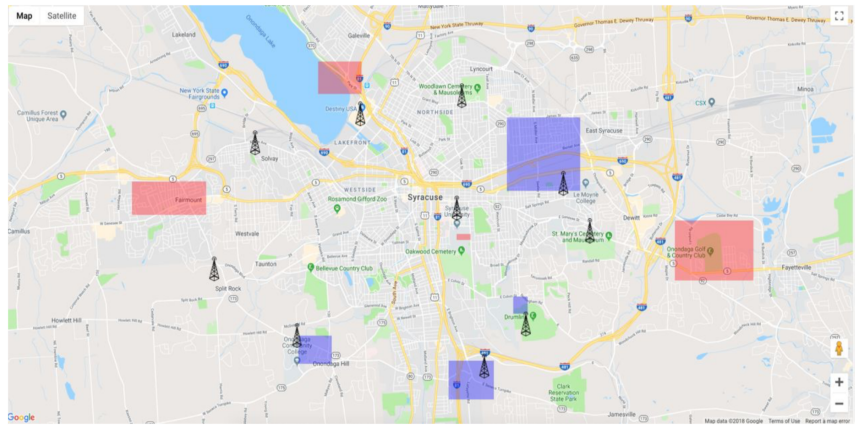

**Figure 2. Configuration Visualization of one Sample Scenario**

By varying the aforementioned configuration parameters, the team generated four air traffic scenarios.

The first scenario is the point to point route without the trajectory management. Nonetheless, other three scenarios follow the Manhattan route. Compared with the second scenario, which is the naive Manhattan style route, the third scenario integrates the trajectory management to avoid potential conflicts. The last scenario introduces the geographical constraint (the no-fly zone) in the simulation.

*Evaluation Metrics*

In the experiments, we applied the MATRUS simulation in each air traffic scenario to measure four metrics: UAS Conflict Ratio, Throughput, Average Flight Time and Average Signal Strength.

To ensure flight safety, two flying sUAS must be separated by a sufficient distance. Because the current sUAS cannot make sharp turns or slow down immediately to a stop, creating an appropriate buffer zone for safety reasons for each sUAS is necessary.

We divided the entire map into multiple square cells. The width of each cell is 18 meters, which is the minimum distance between two sUAS. The total area of each cell is 324 square meters. If a cell is occupied by more than one UAS at the same time, we consider it as a *conflict*.

In our evaluation, the *conflict ratio* is used to analyze the safety metric. It is defined as the number of missions that have encountered at least one conflict during the flight divided by the total number of launched UAV missions.

The UAS *throughput* indicates the capacity of the simulated air space. It is measured by the number of launched sUAS during a fixed time. There is a fundamental tradeoff between safety and throughput. A spatial-temporal trajectory planning algorithm can significantly reduce the conflict ratio, however it will also affect the throughput. The simulation allows us to study the relation between these two.

Besides the overall throughput of the entire simulated air space, the performance of each sUAS is also important. In this paper, we evaluate the *average*

*flight time* of individual sUASs for each scenario and compare the result with the P2P trajectory scenario. Longer average flight time indicates more detour during the flight and higher energy dissipation.

Finally, we evaluated the *signal strength* of the communication channels between the sUAS and base stations. This metric can be used as an indication of the availability and reliability of the C2-link between sUAS and the command center. In the simulation, we choose the log-distance path loss model to to calculate the received signal strength. However, any other channel model can be easily integrated into the simulation by modifying the base station agent.

*Traffic Management for Conflict Elimination*

In the first experiment, we compared the maximum UAS density in different air traffic scenarios, and demonstrate the importance of traffic management. Figures 3,4,5 and 6 visualize the distribution of maximum density for the simulation of each specific scenario. The areas in red rectangles are the sUAS launching areas and the areas in blue rectangles are the landing areas. In the density map, a blue spot indicates normal traffic density, i.e. the maximum density of sUAS in that area is 1 UAS/grid, where each grid is 18x18 $m^2$. In contrast, the bright yellow spot indicates conflict, i.e. the maximum sUAS density is 2 or higher in the specific location. The dark blue areas are where no sUAS have taken place. The detailed comparison results are shown in Table 3, where P2P stands for point to point route and M stands for the Manhattan style route.

The first observation from the first two columns is that the Manhattan style route leads to an increase in the average flight time and conflict ratio. This is because the vertical and horizontal trajectory segments in this type of routing make the routes longer than a straight trajectory. For point to point and Manhattan trajectories, without management, 9.22% and 21.78% missions respectively will have conflicts. On the other hand, with trajectory management, we can eliminate all the conflicts with only a 0.95% increase in the average flight time. The last row shows that even with the geographical constraints (i.e. the no-fly zones), the trajectory management algorithm in the MATRUS framework has the ability to prevent potential flight trajectory conflicts.

**Table 3. Traffic Type Comparison**

| Traffic Type | Avg. Throughput | Avg. Flight Time | Avg. Conflict Ratio |
|---|---|---|---|
| P2P Trajectory w/o Management | 4011 | 381.95s | 9.22% |
| M Trajectory w/o Management | 4008 | 487.64s | 21.78% |
| M Trajectory w. Management | 3832 | 492.25s | 0 |
| M Trajectory with Constraints & Management | 3659 | 522.01s | 0 |

The second observation is that with trajectory management, the average throughput is reduced about 4.4%. And this number becomes 8.7% after incorporating the geographical constraints. This is because with trajectory management, the simulator will cancel the launching of a few sUAS if they are predicted to cause conflicts. Therefore, the more restrictive the constraints are, the fewer sUAS that will be launched.

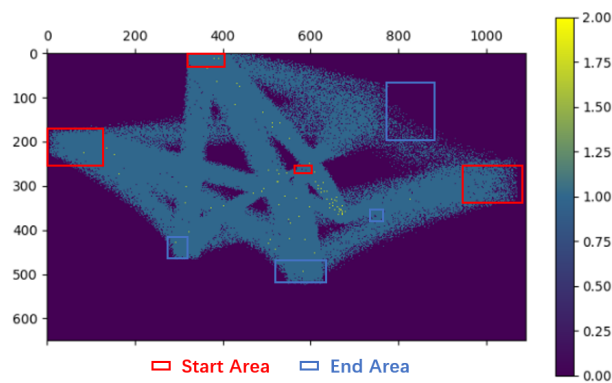

**Figure 3. sUAS Point to Point (P2P) Trajectory Density Map**

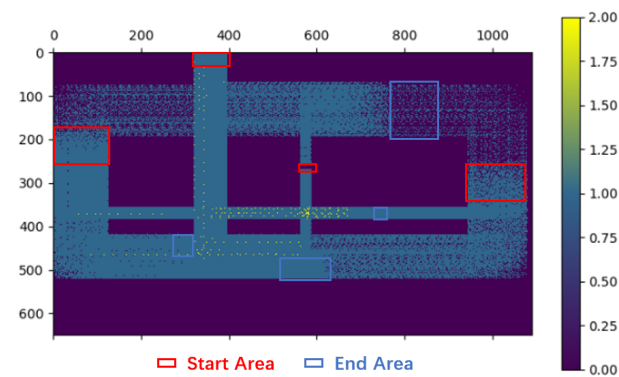

**Figure 4. sUAS Trajectory Density Map without Traffic Management**

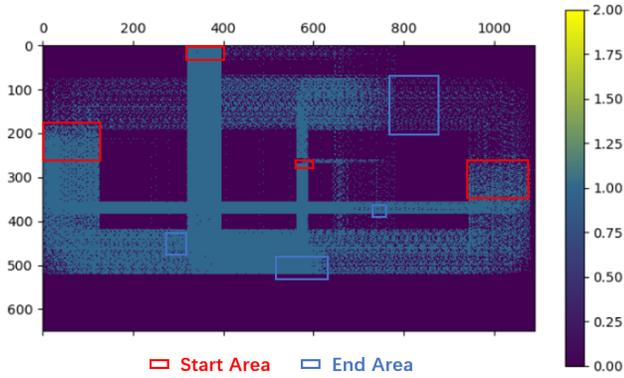

**Figure 5. sUAS Trajectory Density Map with Traffic Management**

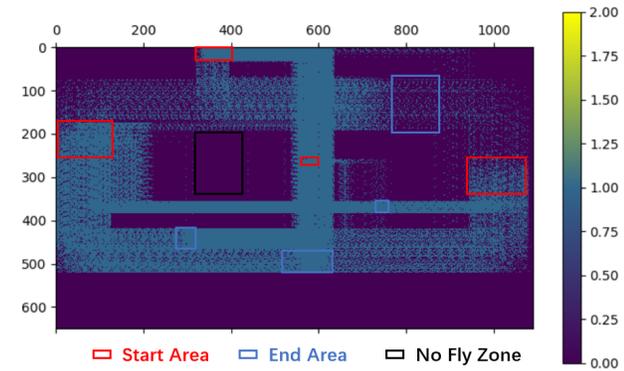

**Figure 6. sUAS Trajectory Density Map with Geographical Constraints and Traffic Management**

## Traffic Management for Improved Connectivity

In our second set of experiments, we compare different air traffic scenarios considering the quality of communication links between sUAS and base stations. The results show that appropriate traffic management can ensure high quality connections between the sUAS and the cellular network.

The distribution (i.e. location) of the base stations impacts the signal strength. The received signal power is calculated based on the log-distance path loss model. If the path loss is less than 80 dB, the link is considered as good quality, and if the loss is greater than 80 dB but less than or equal to 120 dB, then the link is considered as poor quality. Otherwise, if the path loss is greater than 120 dB, the sUAS cannot establish a connection with the base station. Because a link's quality is a function of the sUAS distance to a base station, the coverage map of each base station can be easily visualized in a map. Figure 7 and Figure 8 show the coverage map of good quality and poor quality links respectively.

**Table 4. Signal Strength Comparison**

| Traffic Type | No Link Rate | Poor Link Rate |
|---|---|---|
| P2P Trajectory w/o Management | 0.73% | 44.99% |
| M Trajectory w/o Management | 0.63% | 53.80% |
| M Trajectory with Management | 0.59% | 53.52% |
| M Trajectory with Constraints & Management | 0.58% | 44.94% |

Table 4 shows the distribution of link qualities between sUASs and base stations in different scenarios. From the table we can see that, compared to the straight free-flight, the Manhattan trajectory increases the percentage of poor links from 44.99% to more than 53%, regardless the status of trajectory management. Based on the current base station distribution this occurs because most of the coverage provided by the base stations is gathered in the central area of the map. The Manhattan style trajectory uses vertical and horizontal route path segments in which most of the sUAS pass through the poor link coverage area. Another reason is that the average sUAS flight time is increased. All the sUAS try to use the base station channels with good signal strength. The sUAS that cannot establish a connection with those channels have to connect to the base station with a lower quality channel. The longer average flight time leads to more intensive communication resource competition. The trajectory management does not have any impact on signal quality, because it is not part of the objective function during optimization.

We further notice that by importing the geographical constraints, the percentage of sUAS's that cannot establish a link or can establish only a poor link reduces. This is because the no-fly zone happens to be located outside the area where there is good signal quality as shown in Figure 7. By prohibiting the sUAS from entering this area, we force them to fly into areas that provide good signal quality. This shows that by setting the no-fly zone according to the cellular signal coverage, the trajectory management does not only resolves conflicts, but can also be used to improve the communication quality.

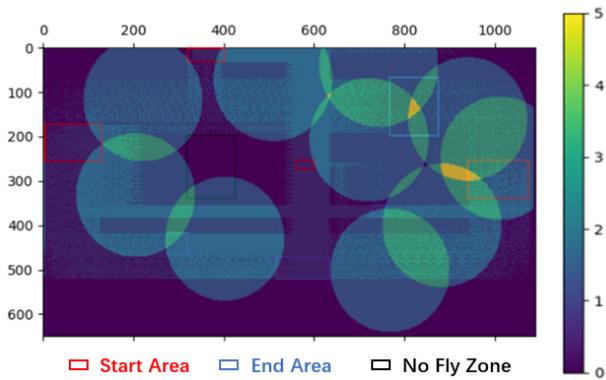

**Figure 7. The Good Signal Quality Coverage Map**

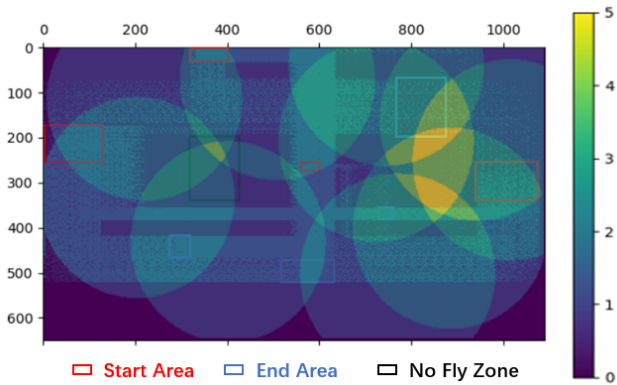

**Figure 8. The Poor Signal Quality Coverage Map**

## Conclusions

In this paper, we have proposed a new Multi-agent Air Traffic and Resource Usage Simulation (MATRUS) framework. The framework is an integrated environment for UTM simulation, data analysis, and traffic visualization. The core functionality of the framework is built on top of an agent-based simulation platform. In addition, various communication and propagation models can be defined in the MATRUS framework and used for evaluating the signal strength between the sUAS and a base station. Our initial experiments have shown that our simulation tool has the ability to evaluate different sUAS traffic management policies. Moreover, the MATRUS framework can provide insights on the relationships between air traffic and communication resource usage for further study.

The UTM concept in urban/suburban environments is still being developed, many technical and operational challenges are still open. In our future work, MATRUS will be instrumental in evaluating UTM traffic management policies, understanding the airspace intrinsic capacity, performing low level airspace traffic optimization. We will also expand the framework to study the effect of sUAS detect and avoid mechanisms, implement additional traffic management policies and handle more complex traffic demand geographical distribution.